\documentclass[onecolumn,pra,showpacs,superscriptaddress]{revtex4}
\usepackage{amssymb}
\usepackage{amsmath}
\usepackage{graphicx}
\usepackage{subfigure}
\usepackage{natbib}
\usepackage{epsfig}
\usepackage{amsfonts}
\usepackage{mathrsfs}
\usepackage{ulem}
\usepackage{color}
\usepackage[toc,page,title,titletoc,header]{appendix}
\begin{document}

\title{Strongly interacting Bose-Fermi mixtures in one dimension}
\author{Haiping Hu}
\affiliation{Beijing National
Laboratory for Condensed Matter Physics, Institute of Physics,
Chinese Academy of Sciences, Beijing 100190, China}

\author{Liming Guan}
\affiliation{National center for nanoscience and technology,
Chinese Academy of Sciences, Beijing 100190, China}

\author{Shu Chen}
\thanks{schen@iphy.ac.cn}
\affiliation{Beijing National
Laboratory for Condensed Matter Physics, Institute of Physics,
Chinese Academy of Sciences, Beijing 100190, China}
\affiliation{Collaborative Innovation Center of Quantum Matter, Beijing, China}

\begin{abstract}

We study one-dimensional strongly interacting Bose-Fermi mixtures by both the exact Bethe-ansatz method and variational perturbation theory within the degenerate ground state subspace of the system in the infinitely repulsive limit. Based on the exact solution of the one-dimensional Bose-Fermi gas with equal boson-boson and boson-fermion interaction strengths, we demonstrate that the ground state energy is degenerate for different Bose-Fermi configurations and the degeneracy is lifted when the interaction deviates the infinitely interacting limit. We then show that the ground properties in the strongly interacting regime can be well characterized by using the variational perturbation method within the degenerate ground state subspace, which can be applied to deal with more general cases with anisotropic interactions and in external traps. Our results indicate that the total ground-state density profile in the strongly repulsive regime behaves like the polarized noninteracting fermions, whereas the density distributions of bosons and fermions display different properties
for different Bose-Fermi configurations and are sensitive to the anisotropy of interactions.
\end{abstract}


\maketitle


\section{Introduction}
One-dimensional (1D) quantum gases have attracted renewed attention during the past decades due to the experimental progress in trapping and manipulating cold atomic systems \cite{RMP-Bose,RMP-Guan}. By loading bosonic or (and) fermioic cold atoms in 1D waveguides, one may realize bosonic or fermionic gases (the Bose-Fermi mixtures). In comparison with the bosonic and fermionc gases, the Bose-Fermi mixtures are particularly
interesting as they rarely occur in nature but are accessible in current cold atomic experiments \cite{Hulet,BF_exp,Fukuhara}.
Theoretical investigations have unveiled the Bose-Fermi mixtures displaying rich phase diagrams and interesting excitation properties \cite{Das,Cazalilla,Mathey,Lai,Imambekov,Yin,Yin2012,Batchelor,Frahm}. The Bose-Fermi mixtures also provide a platform to realize and study physical properties of the Bose-Fermi supersymmetry \cite{Yu,Yang}.

The tunability of interaction strengths between ultracold atoms has
provided unprecedented opportunities for investigating intriguing many-body physics in 1D quantum gases in
the entire parameter regime  \cite
{Paredes,Kinoshita,Olshanii,Gorlitz,Moritz}, with the symbolic
experimental progresses in the realization of Tonks-Girardeau (TG) gas \cite{Paredes,Kinoshita} and super-Tonks-Girardeau (STG) gas \cite{Haller}. Furthermore, recent experiments on few-particle atomic systems with the controllability of precise atom numbers open access to studying few-body physics and the size-dependent evolution from few-body to many-body systems \cite{Jochim1,Jochim2}. Besides the bosonic TG gases, the fermionization of the interacting fermion system has also been observed in the few-particle fermion system \cite{Jochim3}.

It is well known that the TG gas corresponds to a 1D Bose gas with infinitely repulsive interactions between bosonic particles, which was well understood by using the Bose-Fermi mapping proposed by Girardeau in his seminal work more than fifty years ago \cite{Girardeau}. Motivated by the cold atomic experimental progress, the Bose-Fermi mapping was also generalized to study 1D multi-component quantum gases in the strongly repulsive limit \cite{Girardeau07,Deuretzbacher,Guan09,Fang}. Different from the single-component bosonic TG gas, the ground state of multi-component quantum gases in the infinitely repulsive limit is highly degenerate with the degree of degeneracy given by the number of distinct species (spin) configurations \cite{Girardeau07,Deuretzbacher,Guan09,Fang,Guan10,Cui,Harshman}. Slightly away from the infinitely repulsive limit,
a perturbation theory within the degenerate subspace can be constructed using with the inverse of the interaction strength as a small parameter \cite{Guan-thesis,Zinner14,Zinner-SR,PuHan,Deuretzbacher14,Zinner15}.
Particularly, the effective spin-exchange Hamiltonian, which describes the spin dynamics of the spin-1/2 boson and fermion systems in the strongly repulsive regime, has been derived \cite{PuHan,Deuretzbacher14,Zinner15,Levinsen,Zinner-SR}. The trapped multicomponent systems in the full interaction regime have also been studied by exact solutions \cite{Hao-BB,Patu}, highly accurate numerical diagonalization methods \cite{Hao,Blume,Schmelcher,Lewenstein,Bugnion} and variational methods \cite{Zinner-EPL,Barfknecht,Brouzos,Polls,Brouzos2013}.

While most of the previous works concentrated on the strongly interacting bosonic and fermioic systems \cite{Girardeau07,Deuretzbacher,Guan09,Guan10,Guan-thesis,Zinner14,Cui,Zinner-SR,PuHan,Deuretzbacher14,Zinner15,Hao,Blume,Schmelcher,Lewenstein,Harshman,Hao-BB,Patu,Levinsen}, the Bose-Fermi mixtures in the strongly interacting regime are not well studied except for the limiting case with infinite repulsion \cite{Girardeau07,Fang}. It is still not clear how the degenerate ground states split when the interaction deviates the infinite repulsion limit. Another interesting question is how the anisotropy of boson-boson interaction and boson-fermion interaction affects the physical properties of the strongly interacting mixture system? Aiming to answering the above questions, in this work we shall focus our study on the ground state properties of the 1D Bose-Fermi mixtures in the strongly repulsive regime. In order to get some exact results and provide a touchstone for the following results based on the perturbation theory within the degenerate subspace, we first consider the isotropic case with equal boson-boson and boson-fermion interactions, which is exactly solvable by the Bethe-ansatz method in the absence of external potentials. Next we derive the universal energy relation for the Bose-Fermi mixtures and then present our perturbation theory within the degenerate subspace.
By comparing the variational result with the exact result, we find that they agree very well in the strongly repulsive regime. The variational perturbation theory is then applied to deal with the anisotropic mixtures with different boson-boson and boson-fermion interaction strengths trapped in a harmonic trap. Our results indicate that the anisotropic parameter has a significant effect on the ground state density distribution of the Bose and Fermi components, although it almost does not affect the total density distribution of the mixtures in the strongly repulsive regime.

\section{Model and exact solutions for isotropic mixtures}
We consider the 1D interacting Bose-Fermi mixtures described by the Hamiltonian
\begin{eqnarray}
H=\int_0^L dx \{\Psi_b^{\dag}(-\frac{\hbar^2}{2m_b}\partial_x^2+V_b(x))\Psi_b+\Psi_f^{\dag}(-\frac{\hbar^2}{2m_f}\partial_x^2+V_f(x))\Psi_f+
\frac{1}{2}g_{bb}\Psi_b^{\dag}\Psi_b^{\dag}\Psi_b\Psi_b+g_{bf}\Psi_b^{\dag}\Psi_f^{\dag}\Psi_f\Psi_b\} . \label{H}
\end{eqnarray}
Here, $\Psi_b$ and $\Psi_f$ denote the bosonic and fermionic annihilation operators, respectively. The mixtures are confined in external traps $V_b(x)$ and $V_f(x)$. The boson and boson or boson and fermion are interacting through the contact interaction with different strengths $g_{bb}$ and $g_{bf}$ (define anisotropy $\eta=g_{bf}/g_{bb}$), while the interaction between fermions is forbidden by the Pauli exclusion principle.
The model (\ref{H}) in the absence of external potential, i.e., with $V_b(x)=V_f(x)=0$, is exactly solvable, when  $m_b = m_f \equiv m$ and $g_{bb}=g_{bf}\equiv g$ \cite{Lai,Imambekov,Korepin-book}.
While the first condition is approximately fulfilled for isotopes of atoms (for example $^{171}$Yb and $^{172}$Yb, and $^{86}$Rb and $^{87}$Rb), the second condition can be realized by tuning the interaction strength via Feshbach resonances. In this work, we only consider the case with $m_b = m_f \equiv m$, but relax the restriction $g_{bb}=g_{bf}$ when we discuss the anisotropic case with $\eta \neq 1$ by using the variational perturbation theory. Few-body systems with different inter-component and intra-component interaction strengths have been studied in some recent works \cite{Barfknecht,Brouzos,Polls,Brouzos2013,Zinner-EPL}.

Next we shall consider the exactly solvable model with equal masses and equal repulsive boson-boson and boson-fermion interaction strengths. To keep consistent with the traditional references for the integrable Bose-Fermi model, we set $c= m g /\hbar^2$. Consider the Hilbert space spanned by $N$ particles, then the eigenvalue problem reduces to solve the Schr\"{o}dinger equation with first quantized Hamiltonian
\begin{eqnarray}
H=-\sum_i\frac{\partial^2}{\partial x_i^2}+2c\sum_{i<j}\delta(x_i-x_j) .
\end{eqnarray}
Among the $N$ particles, there are $N_b$ bosons and the rest of them are fermions. Under periodic boundary condition, the Bethe-ansatz equations (BAEs) are given by
\begin{eqnarray}
k_j L=2\pi I_j-2\sum_{\alpha=1}^{M}\theta(k_j-\Lambda_{\alpha}),  j=1,...,N\\
2\pi J_{\alpha}=2\sum_{j=1}^{N}\theta(\Lambda_{\alpha}-k_j),    \alpha=1,...M
\end{eqnarray}
where $M=N_b$. We have set $\hbar=2m=1$ and $\theta(x)=\tan^{-1}\frac{x}{c/2}$. The quantum numbers $I_j$ and $J_{\alpha}$ are integers or half integers, depending on the parity of $N_b$ and $N$. In general, $k_j$s are called as quasi-momenta, and $\Lambda_{\alpha}$s are called as rapidities. For $c>0$, both $k_j$s and $\Lambda_{\alpha}$s take real solutions. The eigenenergies are given by $E=\sum_j\frac{\hbar^2}{2m}k_j^2$.

In the strongly repulsive regime ($cL/N\gg1$), $\Lambda_{\alpha}$s are proportional to $c$ while $k_j$s remain finite. The BAEs can be rewritten
in order of $k_j/c$ by using the Taylor expansion
\begin{eqnarray*}
k_jL = 2\pi I_j+\sum_{\alpha =1}^M[2\theta (\Lambda _\alpha )-\frac
4{1+4(\Lambda _\alpha /c)^2}\frac{k_j}c +\frac{16}{(1+4(\Lambda _\alpha /c)^2)^2}\frac{\Lambda _\alpha }c(\frac{k_j}%
c)^2+\cdot \cdot \cdot ] ,
\end{eqnarray*}
and
\begin{eqnarray*}
2\pi J_{\alpha}= \sum_{j=1}^{N}[2\theta (\Lambda _\alpha )-\frac
4{1+4(\Lambda _\alpha /c)^2}\frac{k_j}c +\frac{16}{(1+4(\Lambda _\alpha /c)^2)^2}\frac{\Lambda _\alpha }c(\frac{k_j}%
c)^2+\cdot \cdot \cdot ] ,
\end{eqnarray*}
where we have used $\tan^{-1}(-x)=-\tan^{-1}(x)$ and $d(\tan^{-1}(x))/dx = 1/(1+x^2)$.
Keeping to the first order of $k_j/c$, the quasi-momenta can be calculated:
\begin{eqnarray}
k_j L=2\pi I_j-\zeta\frac{k_j}{|c|}+O((|c|L)^{-3}) ,
\end{eqnarray}
where
\begin{equation}
\zeta=\sum_{\alpha=1}^{M}\frac{1}{(\Lambda_{\alpha}/c)^2+1/4} .
\end{equation}
Here we have used relations
$\sum_{j}k_{j}=0$ and $\sum_{\alpha}\theta(\Lambda_{\alpha})=0$, which are true if $I_j$ and $J_\alpha$ are symmetrical and are always fulfilled
in the case of ground state. By using $2\pi J_{\alpha}= 2 N \theta (\Lambda _\alpha )$, which is just the second BAE under the first order Taylor expansion, we have $\zeta\approx\sum_{\alpha=1}^{M}\frac{4}{\tan^2(\pi J_{\alpha}/N)+1}$. In thermodynamical limit, the sum can be calculated via integration and we have $\zeta=2 M+\frac{2 N}{\pi}sin(\frac{M\pi}{N})$. The energy of the mixture gas in the strongly repulsive limit is thus given by
\begin{eqnarray}
E_{TG}^{BF}=\frac{\hbar^2}{2m}\frac{\pi^2}{3L^2}N(N^2-1)(1+\frac{\zeta}{L|c|})^{-2}+O(|c|^{-3}) .
\end{eqnarray}
In the limit $c\rightarrow \infty$, the ground energy is identical
to that of a polarized N-fermion system and thus is degenerate for different Bose-Fermi configurations. When $c$ deviates the infinitely repulsive limit, the degeneracy is lifted as the value of $\zeta$ is dependent on the number of bosons. We note that
$\zeta(M_1) \leq \zeta(M_2) $ if $M_1 \leq M_2$, which leads to $E_{TG}^{BF} (N_{b1}) \geq E_{TG}^{BF} (N_{b2})$ when $N_{b1} \leq N_{b2}$ for systems with fixed total particle numbers.
\begin{figure}
\includegraphics[width=6in]{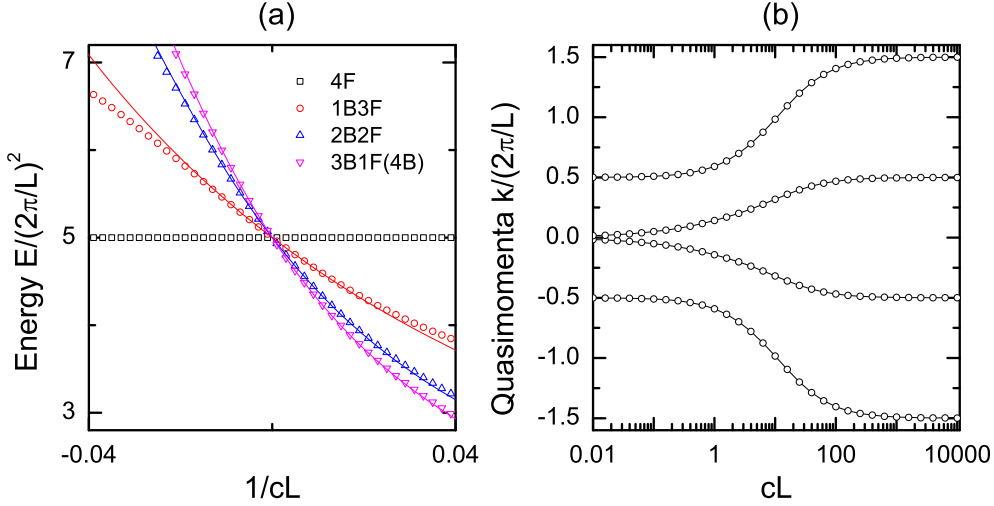}
\caption{(Color online) (a) Ground state energy ($c>0$) and the lowest energy of the scattering state ($c<0$) calculated by BAEs for mixtures with the fixed particle number $N=4$ but different Bose-Fermi configurations. The solid lines represent results obtained from the Taylor expansion. (b) The quasi-momenta distribution versus the interaction strength for the ground state of the $2B2F$ system.}
\end{figure}

For attractive interactions with $c<0$, the BAEs can still have real solutions, which describe the scattering states of the attractive systems. The lowest states with real solutions in the strongly attractive regime correspond to the STG state, similar to STG states in the attractive Bose systems \cite{STG,STG1,Chen10,Hao11} and Fermi systems \cite{Guan10}. We have
\begin{eqnarray}
E_{STG}^{BF}=\frac{\hbar^2}{2m}\frac{\pi^2}{3L^2}N(N^2-1)(1-\frac{\zeta}{L|c|})^{-2}+O(|c|^{-3}) .
\end{eqnarray}
In the limit $c\rightarrow -\infty$,, we have $E_{STG}^{BF}=E_{TG}^{BF}$. The degeneracy is also lifted when $c$ deviates the infinitely attractive limit, and we have $E_{STG}^{BF} (N_{b1}) \leq E_{STG}^{BF} (N_{b2})$ when $N_{b1} \leq N_{b2}$.

To give concrete examples, in Fig.1(a), we demonstrate the ground state energies versus the inversion of interaction strength $1/c$ for Bose-Fermi mixtures with total particle number $N=4$. Totally there are five cases with different Bose-Fermi configurations: $4B$, $3B1F$, $2B2F$, $1B3F$ and $4F$, where the configuration $nBmF$ represents the system composed of $n$ bosons and $m$ fermions. These different states are degenerate at the exact Girardeau's mapping point $c \rightarrow \infty $ with energy $E_{\infty}=\frac{5\hbar^2}{2m}(\frac{2\pi}{L})^2$. The first two cases make no difference since no Fermi statistic should be considered. We can clearly see that the degeneracy of the ground state energy is lifted when the system deviates the exact mapping point. At the repulsive side, i.e., in the TG regime, the ground state energy decreases with the decrease of $c$. Systems with more bosons have relative lower energies at this regime. While at the attractive side, i.e., in the STG regime, the energy increases with the decrease of $|c|$, and systems with more bosons have relative higher energies. To the linear order, they are perfectly described by the above expansion formula from the BAEs. To be more intuitive, Fig.1(b) shows the distribution of quasi-momenta with respect to the interaction strength for the $2B2F$ system in the whole repulsive regime. At the non-interacting limit, $c=0$, the quasi-momenta should be $0$ for bosons and $\pm\frac{\pi}{L}$ for fermions. As the interaction increases, the quasi-momenta distribution becomes wider and finally at $c= \infty$, we have $k_j=2\pi I_j/L$. The whole system behaves like a polarized fermion system with quasi-momenta $\pm\frac{3\pi}{L}$, $\pm\frac{\pi}{L}$, which is consistent with the result by the generalized Bose-Fermi mapping \cite{Girardeau07}.

\section{Universal energy relation for the 1D Bose-Fermi mixture}
Next we shall derive the universal energy relation for the 1D Bose-Fermi mixtures, which can be viewed as a generalization of Tan's universal energy relation for the spin-1/2 Fermi gas \cite{Guan-thesis,Tan,Braaten,1dTan}. For convenience, we only consider the case with $m_a=m_b=m$ and $V_a(x)=V_b(x)=V(x)$ and the Hamiltonian can be rewritten as the following form
\begin{eqnarray}
H=-\sum_{i} [\frac{\hbar^2}{2 m}\frac{\partial^2}{\partial x_{i}^2}+V(x_{i})] + g_{bb} \sum_{i<j} \delta(x_{i}-x_{j}) \delta^{b}_{\sigma_i, \sigma_j} + g_{bf}\sum_{i<j}\delta(x_{i}-x_{j}) \delta_{\sigma_i, - \sigma_j} ,\label{H-contact}
\end{eqnarray}
where $\sigma_i=b,~f$ represent the bosonic and fermionic components, respectively,  $\delta^{b}_{\sigma_i, \sigma_j}= 1$ only if $\sigma_i$=$\sigma_j$=$b$ and
$\delta_{\sigma_i, - \sigma_j}=1$ when ($\sigma_i=b$,  $\sigma_j=f$) or ($\sigma_i=f$,  $\sigma_j=b$). Here the summation to the spin index is assumed. Let $\Psi(x_1,\sigma_1; x_2,\sigma_2;...;x_N,\sigma_N)$ be the normalized eigenstate of the system, which fulfills the Schrodinger equation:
\begin{eqnarray}
H\Psi(x_1,\sigma_1; x_2,\sigma_2;...;x_N,\sigma_N)=E \Psi(x_1,\sigma_1; x_2,\sigma_2;...;x_N,\sigma_N) .
\end{eqnarray}
Denote $x_1,x_2$ as the coordinates of two interacting particle with interaction $g_{bb}$ or $g_{bf}$, depending on whether they are two bosons or one boson and one fermion. Suppose that $x_3 , .., x_N$ do not coincide with $x_1$ and $x_2$. In terms of center-of-mass and relative coordinates, $x=(x_1+x_2)/2$ and $r=x_1-x_2$, integrating $r$ around $0$, one gets
\begin{eqnarray}
\frac{\partial\Psi}{\partial r}|_{r=0^+}-\frac{\partial\Psi}{\partial r}|_{r=0^-}=\frac{m g_{bb}}{\hbar^2} \delta^{b}_{\sigma_1, \sigma_2} \Psi|_{r=0}+\frac{m g_{bf}}{\hbar^2}\delta_{\sigma_1, - \sigma_2} \Psi|_{r=0} .
\end{eqnarray}
The equation can be rewritten as
\begin{eqnarray}
\frac{\partial\Psi}{\partial x_1}|_{x_1=x_2^+}-\frac{\partial\Psi}{\partial x_1}|_{x_1=x_2^-}+
\frac{\partial\Psi}{\partial x_2}|_{x_2=x_1^+}-\frac{\partial\Psi}{\partial x_2}|_{x_2=x_1^-}=
\frac{2 m g_{bb}}{\hbar^2} \delta^{b}_{\sigma_1, \sigma_2} \Psi|_{r=0}+\frac{2 m g_{bf}}{\hbar^2}\delta_{\sigma_1, - \sigma_2} \Psi|_{r=0} .
\end{eqnarray}
This is just the boundary condition required for two interacting particles. At small $r$, we have the Taylor expansion \cite{Guan-thesis}:
\begin{eqnarray}
& & \Psi(x_1=x-r/2,\sigma_1;x_2=x+r/2,\sigma_2;X) \notag \\
&=& \delta^{b}_{\sigma_i, \sigma_j} [
A^{bb}(x,X)(|r|-a_{bb})+B^{bb}(x,X)r]+\delta_{\sigma_i, -\sigma_j} [
A^{bf}(x,X)(|r|-a_{bf})+B^{bf}(x,X)r]+O(r^2) ,
\end{eqnarray}
where $X=(x_3,\sigma_3;...x_N,\sigma_N)$, $a_{bb}=-\frac{2\hbar^2}{m g_{bb}}$ and $a_{bf}=-\frac{2\hbar^2}{m g_{bf}}$ are the effective 1D scattering lengths,
\begin{eqnarray}
A^{bb} (x,X)&=& \frac{1}{2}\left[ \frac{\partial\Psi(x_1,b;x_2,b;X)}{\partial x_1}|_{x_1=x_2^+}-\frac{\partial\Psi(x_1,b;x_2,b;X)}{\partial x_1}|_{x_1=x_2^-} \right]=-\frac{1}{a_{bb}}\Psi(x,b;x,b;X) ,\\
A^{bf} (x,X)&=& \frac{1}{2}\left[ \frac{\partial\Psi(x_1,\sigma;x_2,-\sigma;X)}{\partial x_1}|_{x_1=x_2^+}-\frac{\partial\Psi(x_1,\sigma;x_2,-\sigma;X)}{\partial x_1}|_{x_1=x_2^-} \right] =-\frac{1}{a_{bf}}\Psi(x,\sigma;x,-\sigma;X) ,
\end{eqnarray}
and
\begin{eqnarray*}
B^{bb}(x,X) &=& \frac{1}{2}\left[ \frac{\partial\Psi(x_1,b;x_2,b;X)}{\partial x_2}|_{x_1=x,x_2=x^-}-\frac{\partial\Psi(x_1,b;x_2,b;X) }{\partial x_1}|_{x_1=x^-,x_2=x} \right], \\
B^{bf}(x,X) &=& \frac{1}{2}\left[ \frac{\partial\Psi(x_1,\sigma;x_2,-\sigma;X)}{\partial x_2}|_{x_1=x,x_2=x^-}-\frac{\partial\Psi(x_1,\sigma;x_2,-\sigma;X)}{\partial x_1}|_{x_1=x^-,x_2=x} \right] .
\end{eqnarray*}
Similarly, for a different coupling $g'$ with energy $E'$, we have
\begin{eqnarray}
& & \Psi'(x_1=x-r/2,x_2=x+r/2;X) \notag \\
&=& \delta^{b}_{\sigma_i, \sigma_j} [
A'^{bb}(x,X)(|r|-a'_{bb})+B'^{bb}(x,X)r]+\delta_{\sigma_i, -\sigma_j} [
A'^{bf}(x,X)(|r|-a'_{bf})+B'^{bf}(x,X)r]+O(r^2) .
\end{eqnarray}

From $H\Psi=E\Psi$ and $H'\Psi'=E'\Psi'$, we get $\Psi'^{*}H\Psi-\Psi'^{*}H'\Psi=(E-E')\Psi'^{*}\Psi$. Integrating this equation over the coordinates and summing over all the interacting pairs, we have
\begin{eqnarray}
&& \mathcal{N}^{bb}_P \frac{4\hbar^2}{m^2}(\frac{1}{g_{bb}'}-\frac{1}{g_{bb}})\int dxdX A_{bb}'^*(x,X)A_{bb}(x,X)+\mathcal{N}^{bf}_P\frac{4\hbar^2}{m^2}(\frac{1}{g_{bf}'}-\frac{1}{g_{bf}})\int dxdX A_{bf}'^*(x,X)A_{bf}(x,X) \notag \\
&=& -(E'-E)\int dx_1...dx_N \Psi'^*\Psi ,
\end{eqnarray}
where $\int dX=\int dx_3...dx_N$, $\mathcal{N}^{bb}_P=N_b(N_b-1)/2$ for boson-boson interactions and $\mathcal{N}^{bf}_P=N_b N_f$ for boson-fermion interactions. Taking the limit of $g'\rightarrow g$ and $\Psi'\rightarrow \Psi$, we get
\begin{eqnarray}
{d E} = {- d (1/g_{bb})} \frac{4\hbar^4}{m^2} I_{bb} - {d (1/g_{bf})} \frac{4\hbar^4}{m^2} I_{bf},
\label{energy}
\end{eqnarray}
where $I_{bb}$ and  $I_{bf}$ are the contacts defined by
\begin{eqnarray}
I_{bb}=\mathcal{N}_P^{bb}\int dxdX|A^{bb}(x,X)|^2, \\
I_{bf}=\mathcal{N}_P^{bf}\int dxdX|A^{bf}(x,X)|^2.
\end{eqnarray}
From Eq.(\ref{energy}), we have
\begin{eqnarray}
\frac{\partial E}{\partial (-1/g_{bb})}=\frac{4\hbar^4}{m^2} I_{bb}, ~~~~~ \frac{\partial E}{\partial (-1/g_{bf})}=\frac{4\hbar^4}{m^2} I_{bf}.
\end{eqnarray}
The above relations are the universal energy relations of the Bose-Fermi mixtures.

For the Bose-Fermi mixtures with fixed anisotropy parameter $\eta$, we can get the expression for the energy near the infinitely interacting limit
\begin{eqnarray}
E=E_{\infty}-\frac{1}{g_{bb}}\frac{4\hbar^4}{m^2}(I_{bb}+\frac{I_{bf}}{\eta}),
\end{eqnarray}
where $E_{\infty}$ represents the energy of the system at the TG limit.
Some results directly follow from this formula. First, since $I_{bb}\propto N_b(N_b-1)/2$ and $I_{bf}\propto N_b N_f$, for the isotropic interacting case with $\eta=1$ and fixed particle numbers $N$, we can conclude that systems with more bosons have lower energies. When the boson-boson interaction dominates, $\eta\ll 1$, the second term in the contact matrix requires the system to have the lowest energy at $N_b=N_f$. When the boson-fermion interaction dominates, $\eta\gg 1$, the first term requires the state with more bosons having lower energy. Generally, for $\eta<1$, the states with the relative ratio $N_b/N\propto \frac{1}{2-\eta}$ have the lowest energy. Our universal relations derived in this section coincide with the exact results given by the BAEs.

\section{Variational perturbation method}
The Bethe-ansatz method is powerful but only limited to the integrable case with $g_{bb}=g_{bf}$ and $V(x)=0$. While for a trapped system or the system with anisotropic interactions $g_{bb} \neq g_{bf}$, we need develop a more general method to address the problem. We notice that there exists one exact soluble point $1/g=0$, i.e., the infinitely repulsive limit $g_{bb}=g_{bf}= \infty$, in which the system can be mapped to the polarized fermionic system and the many-body wave function can be constructed from the single-particle fermionic wave function, taking into account the statistics of exchange between bosons or fermions \cite{Girardeau07,Fang}. This is the central idea of the generalization of Girardeau's Bose-Fermi mapping to the multicomponent systems \cite{Girardeau07,Fang,Deuretzbacher,Guan09}. Since no symmetry is required for the exchange between bosonic and fermionc particles, there exists degeneracy for the eigenenergy. For the system composed of $N_b$ bosons and $N_f$ spinless fermions, the degeneracy is $D=\frac{(N_b+N_f)!}{N_b! N_f!}$, which corresponds to different configurations of $N_b$ bosons in $N$ single-particle states.

Once the system deviates from the infinite repulsion limit, the degeneracy of ground state manifold is lifted. As long as the system is still in the strongly interacting regime with the inverse of interaction strengths much smaller than the single particle level spacing, we can utilize the degenerative perturbation to determine energy splitting.
For our system, at infinite interaction strength, the ground state is D-fold degenerate and the energy can be expressed as $E=\sum_{l=1}^{N}\epsilon_{j_l}$ where $j_l$s are N different integers and $\epsilon_{j_l}$ is the $j_l$-th single particle energy level in the trapping potential $V(x)$ with corresponding wave function $\phi_{j_l}(x)$. In this paper, we focus on the ground state where $j_l=l$. We should stress that our method can also be applied to excited states which depend on the $j_l$ we choose. By the generalized Bose-Fermi mapping \cite{Girardeau07,Deuretzbacher,Guan09,Fang}, the coordinate part of the many-body wave function can be constructed from the anti-symmetric Slater determinant
\begin{eqnarray}
\psi_A(x_1,x_2,...,x_N)=\frac{1}{\sqrt{N!}}\sum_{P} sgn(P) \prod\phi_{P_j}(x_j),
\end{eqnarray}
where $P$ is a permutation of the integers $(1,2,...N)$ and $sgn(P)=\pm 1$ for even and odd permutations. For the infinite interaction strengths the particles can not penetrate each other and the real space can be divided into $N!$ distinct parts. Introducing
$$ \theta_{\alpha}=\left\{
\begin{aligned}
1 &   &(x_{\alpha_1}<x_{\alpha_2}<...<x_{\alpha_N}) \\
0 &   &(others) \\
\end{aligned}
\right.
$$
where $\alpha$ is a sequence of $[1,2,...N]$. Totally we can construct $N!$ orthogonal basis $\psi_A \theta_{\alpha}$. Once we consider constrains of the Bose-Fermi statistics, that is, exchange between fermions should contribute a minus sign while no sign will emerge for the exchange between bosons, the number of allowed eigenfunctions is reduced to $D=\frac{(N_b+N_f)!}{N_b! N_f!}$, which gives the degeneracy of the considered state. So these constructed states can be used as the basis of the degenerate space. Define permutation operators for bosons and fermions as $P_b$ and $P_f$, respectively, we can get the $D$ normalized and orthogonal basis of the degenerate subspace as follows:
\begin{eqnarray}
\psi_{\alpha}(x_1,x_2,...,x_N)=\sqrt{D}\sum_{P_b,P_f}(-1)^{P_b} (P_b P_f \theta_{\alpha})\psi_A ,
\end{eqnarray}
where $(-1)^{P_b}=\pm 1$ for even and odd permutations between bosons.
For the case with a finite large interaction, the eigenfunction should approach the infinite one smoothly when $g$ goes to infinity and it therefore should go into the degenerate subspace. Introducing the projection operator:
 $P_{deg}=\sum_{\alpha}|\psi_{\alpha}\rangle\langle\psi_{\alpha}|$,
then it is reasonable to expand the eigenfunction at finite interaction as
\begin{eqnarray}
\Psi(g,x_1,...,x_N)=\sum_{\alpha}a_{\alpha}\psi_{\alpha} ,
\end{eqnarray}
with $\sum_{\alpha}|a_{\alpha}^2|=1$.
The Bose-Bose contact $I_{bb}$ and Bose-Fermi contact $I_{bf}$ can be calculated using the reduced contact matrices $J^{bb}$ and $J^{bf}$ as:
\begin{eqnarray}
I_{bb}=\sum_{\alpha,\alpha'}a_{\alpha}^* a_{\alpha'}J^{bb}_{\alpha,\alpha'}=\overrightarrow{a}J^{bb}\overrightarrow{a}', \\
I_{bf}=\sum_{\alpha,\alpha'}a_{\alpha}^* a_{\alpha'}J^{bf}_{\alpha,\alpha'}=\overrightarrow{a}J^{bf}\overrightarrow{a}',
\end{eqnarray}
where $\overrightarrow{a}=(a_1,a_2...a_D)^T$ and the reduced matrices for Bose-Bose and Bose-Fermi interaction are defined as
\begin{eqnarray}
J^{bb}_{\alpha,\alpha'}= \mathcal{N}^{bb}_p\int A^{bb *}_{\alpha}(x,X) A^{bb}_{\alpha'}(x,X)dx dX ,\\
J^{bf}_{\alpha,\alpha'}=  \mathcal{N}^{bf}_p\int A^{bf *}_{\alpha}(x,X) A^{bf}_{\alpha'}(x,X)dx dX .
\end{eqnarray}
Then the energy can be read as:
\begin{eqnarray}
E=E_{\infty}-\frac{1}{g_{bb}}\frac{4\hbar^4}{m^2}\sum_{\alpha,\alpha'}a_{\alpha}^* a_{\alpha'}[J^{bb}_{\alpha,\alpha'}+\eta^{-1}J^{bf}_{\alpha,\alpha'}] .
\end{eqnarray}
Finally, we define total contact as $J=J^{bb}+\eta^{-1}J^{bf}$. This is the key quantity which determines the behavior of our system. We can clearly see that the quantum state is largely dependent on the ratio $\eta$ of the two interaction strengths.

The contact can be determined via the variational principle \cite{Guan-thesis,Zinner14}. Let $L=E-\lambda(\sum_{\alpha}a_{\alpha}^* a_{\alpha}-1)$, the variational principle requires:
\begin{eqnarray}
\sum_{\alpha'}J_{\alpha,\alpha'}a_{\alpha'}=\lambda a_{\alpha} . \label{eq-contact}
\end{eqnarray}
From the above equation we can see $\lambda$ and $\alpha$ are eigenvalue and eigen-vector of the total $D\times D$ contact matrix $J$ \cite{Guan-thesis}. The diagonalization procedure gives the splitting energy of original degenerate manifold:
\begin{eqnarray}
E= E_{\infty}-\frac{4\hbar^4}{m^2}\frac{\lambda}{g_{bb}} .
\end{eqnarray}

To show the validity and power of the method, we compare the results with expansions given by BAEs for $g_{bb}=g_{bf}$ in uniform space with periodic boundary condition. As $1/g=0$, $k_j=2\pi/I_j$, the totally anti-symmetric wave-function is $\Psi_A(x_1,...x_N)=(N!)^{-\frac{1}{2}}L^{-\frac{1}{2}N}i^{\frac{1}{2}(N-1)}\exp[-i(N-1)\pi L^{-1}\sum_j x_j]\prod_{j>l}[\exp(i 2\pi L^{-1}x_j)-\exp(i 2\pi L^{-1}x_l)]$. Take $N=3$ and $N=4$ cases as examples. The eigenvalues for the ground states given by the BAEs to the first order Taylor expansions are $\frac{48\hbar^4\pi^2}{m^2L^3}$ and $\frac{160\hbar^4\pi^2}{m^2L^3}$, respectively. For the variational perturbation method, the largest eigenvalue by the diagonalization of the contact matrix produces the same result for the same system.

\section{Application to few-particle systems in a harmonic trap}

In the above section, we have generally discussed the variational perturbation method and introduced the reduced contact matrix to simplify the strongly interacting problem to the matrix diagonalization. In this section, we apply this method to study few-particle Bose-Fermi mixtures with $N=3$ and $N=4$ in a 1D harmonic trap, i.e., $V(x)=\frac{1}{2}m\omega^2 x^2$. As the trap potential preserves the inversion symmetry, the eigenstates are characterized by distinct parities. Our discussion is valid for finite but strong interaction strengths. The interaction $g$ is larger than any other scale in the problem, i.e., $g/\hbar\omega a_{\omega}\gg1$. The single-particle state can be expressed as $\phi_i(x)=\frac{1}{\pi^{1/4}a_{\omega}^{1/2}\sqrt{2^i i!}}H_{i}(\xi)e^{-1/2\xi^2}$ where $H_i({\xi})$ is the Hermite polynomial and $\xi=x/a_{\omega}\equiv x/\sqrt{\hbar/{m\omega}}$.

For the N-particle system, the Slater determinant composed of the $N$ lowest eigen-functions is $\Delta=C_N [\prod_{i=1}^{N}e^{-\xi_i^2/2}]\prod_{1\leq j<k\leq N}(\xi_k-\xi_j)$ with coefficient $C_N=2^{N(N-1)/4}a_{\omega}^{-N/2}[N!\prod_{n=0}^{N-1}(n!\sqrt{\pi})]^{-1/2}$  \cite{girardeaupra}. We first take the $2B1F$ case as an example. At the infinite repulsion limit, the ground state is $3$-fold degenerate. Denote $x_1$, $x_2$ and $x_3$ as coordinates of bosons and fermions, respectively. We can define the following three distinct subspace basis respecting exchange statistics:
\begin{eqnarray}
\psi_1=\sqrt{3}(\theta(123)-\theta(213))\Delta,\\\notag
\psi_2=\sqrt{3}(\theta(132)-\theta(231))\Delta,\\\notag
\psi_3=\sqrt{3}(\theta(312)-\theta(321))\Delta.
\end{eqnarray}
An explicit calculation gives the contact matrices for the boson-boson interaction and boson-fermion interaction as
\begin{eqnarray}
J^{bb}=\frac{27}{64 \sqrt{2\pi}a_{\omega}^3}\left(
\begin{array}{ccc}
4 & 0 & 0\\
0 & 0 & 0\\
0 & 0 & 4\\
\end{array}
\right)
, ~ J^{bf}=\frac{54}{64 \sqrt{2\pi}a_{\omega}^3}\left(
\begin{array}{ccc}
1  & -1  & 0 \\
-1 &  2  & -1\\
0  & -1  & 1 \\
\end{array}
\right).
\end{eqnarray}
By solving the eigen-equation (\ref{eq-contact}) for the contact matrix, we can directly get the variational wavefunctions and energies. We note that the wavefunctions obtained in our scheme automatically fulfill the symmetry of parity. For example, for the isotropic case with $\eta=1$, we get three eigenvalues of total contact matrix: $\lambda_1=\frac{216}{64 \sqrt{2\pi}a_{\omega}^3}$, $\lambda_2=\frac{162}{64 \sqrt{2\pi}a_{\omega}^3}$, $\lambda_3=\frac{54}{64 \sqrt{2\pi}a_{\omega}^3}$. The corresponding normalized eigenvectors $(a_1,a_2, a_3)^T$ are: $\frac{1}{\sqrt{3}}(-1,1,-1)^T$, $\frac{1}{\sqrt{2}}(-1,0,1)^T$, $\frac{1}{\sqrt{6}}(1,2,1)^T$. It is clear that the ground state and the second excited state have even parity while the first excited state has odd parity as the even parity require $a_1=a_3$ and the odd parity state requires $a_2=0$ and $a_1=-a_3$. Since the interaction terms do not change the parity of the eigenstate, the parity of the many-body state does not change even we tune the interaction strength continuously to the non-interacting limit, and thus the three many-body states can be adiabatically connected to their noninteracting limits \cite{Zinner14}, labeled by $(3,0,0)$,$(2,1,0)$ and $(1,2,0)$, respectively. Here $(n_1,n_2,n_3)$ means that the occupation numbers on the three lowest single-particle states are $n_1$ ,$n_2$ and $n_3$, respectively.
\begin{figure}
\includegraphics[width=7in]{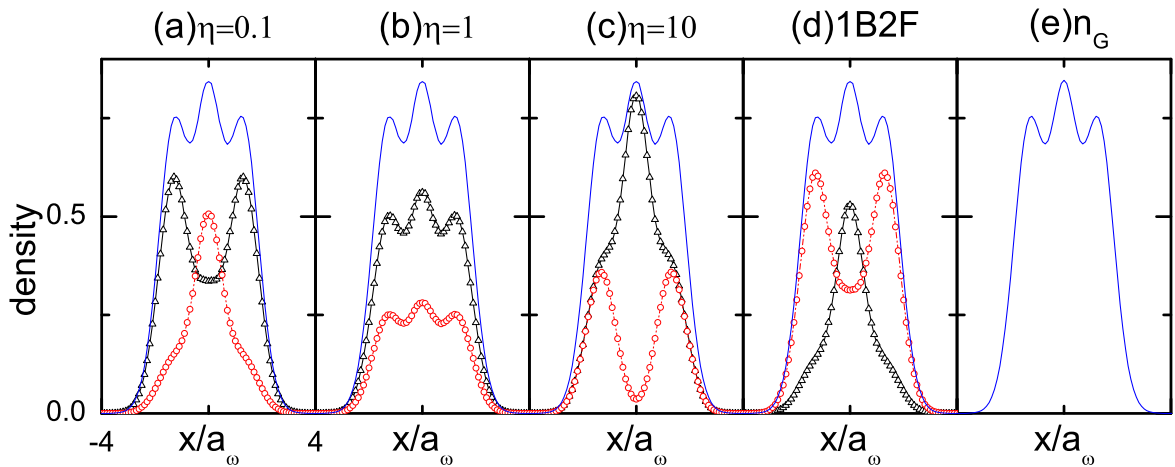}
\caption{(Color online) Ground state density distributions for the $N=3$ mixtures obtained by the variational perturbation method. (a)-(c) are for the $2B1F$ system with different anisotropic interaction ratios $\eta$, i.e., (a) $\eta=0.1$, (b) $\eta=1$ and (c)$\eta=10$. (d) Density distributions for the $1B2F$ case. The red circle, black triangle and blue solid curve represent fermion, bosons and total density distributions, respectively. (e) Total density distribution in the TG limit for $N=3$ mixtures.}
\end{figure}

The density profiles can be calculated from the reduced sing-particle density matrices as:
\begin{eqnarray}
n_b(x,y)&=& N_b\int dX'\Psi^*(x,X')\Psi(y,X') , \\
n_f(x,y)&=& N_f\int dX^{''}\Psi^*(X^{''},x)\Psi(X^{''},y) ,
\end{eqnarray}
where $X'=(x_2,...,x_N)$ and $X^{''}=(x_1,...x_{N-1})$. The diagonal elements are nothing but the single-particle density profiles, i.e., $\rho_b(x)=n_b(x,x)$ and $\rho_f(x)=n_f(x,x)$. In Fig.2 (a)-(c), we show the density distribution for the $2B1F$ system with different interaction anisotropy $\eta$. For $\eta \ll 1$, the boson-boson interaction dominates. As expected, the bosons will be repelled to the two wings while the single fermion mainly locates in the middle regime. For $\eta \gg 1$, the boson-fermion interaction dominates, the overlap between bosons and fermion must be the smallest to avoid the strong inter-species interaction. The bosons can locate on the middle regime and the fermion must be repelled from the center to form two peaks. Particularly, for the case of $\eta =1$, the two interactions equally compete and we have $n_{b}=\frac{2}{3} n_{G}$ and $n_{f}=\frac{1}{3} n_{G}$, where $n_G=\sum_{i}^{N} |\phi_i(x)|^2$ denotes the density distribution in the TG limit. On the other hand, for the $1B2F$ case, the fermions are repelled to the wings with the bosons located at the trap center as shown in Fig.2 (d). In all cases, the totally density distribution is nearly as the same as $n_G$.

Next, we consider the equal-mixing mixtures composed of 2 bosons and 2 fermions in the harmonic trap, i.e., the $2B2F$ case. At the infinitely repulsive limit, the ground state is $6$-fold degenerate. Denote $x_1,x_2$ and $x_3,x_4$ as coordinates of bosons and fermions, respectively. We can define the six distinct subspace basis as follows according to exchange statistics:
\begin{eqnarray}
\psi_1=\sqrt{6}(\theta(1234)-\theta(2134)+\theta(1243)-\theta(2143))\Delta,\\\notag
\psi_2=\sqrt{6}(\theta(1324)-\theta(2314)+\theta(1423)-\theta(2413))\Delta,\\\notag
\psi_3=\sqrt{6}(\theta(1342)-\theta(2341)+\theta(1432)-\theta(2431))\Delta,\\\notag
\psi_4=\sqrt{6}(\theta(3124)-\theta(3214)+\theta(4123)-\theta(4213))\Delta,\\\notag
\psi_5=\sqrt{6}(\theta(3142)-\theta(3241)+\theta(4132)-\theta(4231))\Delta,\\\notag
\psi_6=\sqrt{6}(\theta(3412)-\theta(3421)+\theta(4312)-\theta(4321))\Delta.
\end{eqnarray}
The contact matrices for the boson-boson interaction and boson-fermion interaction are given by
\begin{eqnarray}
J^{bb}=\frac{8}{3\sqrt{\pi}a_{\omega}^3}\left(
\begin{array}{cccccc}
C_1 & 0   & 0   &   0   &  0  &  0\\
0   & 0   & 0   &   0   &  0  &  0\\
0   & 0   & 0   &   0   &  0  &  0\\
0   & 0   & 0   &   C_2 &  0  &  0\\
0   & 0   & 0   &   0   &  0  &  0\\
0   & 0   & 0   &   0   &  0  &  C_1\\
\end{array}
\right), ~
J^{bf}=\frac{1}{3\sqrt{\pi}a_{\omega}^3}\left(
\begin{array}{cccccc}
C_2  & -C_2     & 0     &  0     &  0         &  0 \\
-C_2 & 2C_1+C_2 & -C_1  &  -C_1  &  0         &  0 \\
0    & -C_1     & 2C_1  &  0     &  -C_1      &  0 \\
0    & -C_1     & 0     &  2C_1  &  -C_1      &  0 \\
0    & 0        & -C_1  &  -C_1  &  2C_1+C_2  &  -C_2 \\
0    & 0        & 0     &  0     &  -C_2        &  C_2 \\
\end{array}
\right).
\end{eqnarray}
Here $C_1=0.5938$ and $C_2=0.7796$ are integral constants.
The parity of the state can be directly read out from the eigenvectors of the contact matrix. For the even parity state, $a_1=-a_6$, $a_2=-a_5$ and $a_3=a_4=0$ while for the odd parity state the coefficient should satisfy: $a_1=a_6$, $a_2=a_5$. The diagonalization of the total contact matrix demonstrates that there always exist four odd and two even parity states in the six degenerate subspaces. Similar to $2B1F$ case, we can label these states by their adiabatically connections with the non-interacting states. From the ground state to the fifth excited state, the corresponding occupations in the single-particle orbital are (3,1,0,0), (2,2,0,0), (1,3,0,0), (2,1,1,0), (1,2,1,0) and (1,1,2,0), respectively.

The density profiles are shown in Fig.3 for different interaction ratio $\eta$, exhibiting a bit of difference from the $2B1F$ case. First, when $\eta\ll 1$, the boson-boson interaction dominates and the two bosons behave like hard-core bosons. As shown in Fig.3(a), for the $2B2F$ system with $\eta=0.1$, the density distribution of bosons has almost the same distribution as that of fermions and we have $n_{b}\approx\frac{1}{2} n_{G}$ and $n_{f}\approx\frac{1}{2} n_{G}$. As $\eta$ increases, the boson-fermion interaction gradually dominates and bosons and fermions will repel each other. The fermions are repelled from the harmonic trap center while the bosons will eventually localize at the trap center for $\eta=10$. We notice that while each component in the three different cases has quite different density distribution, the total density distribution is nearly as the same as $n_G$.

\begin{figure}
\includegraphics[width=7in]{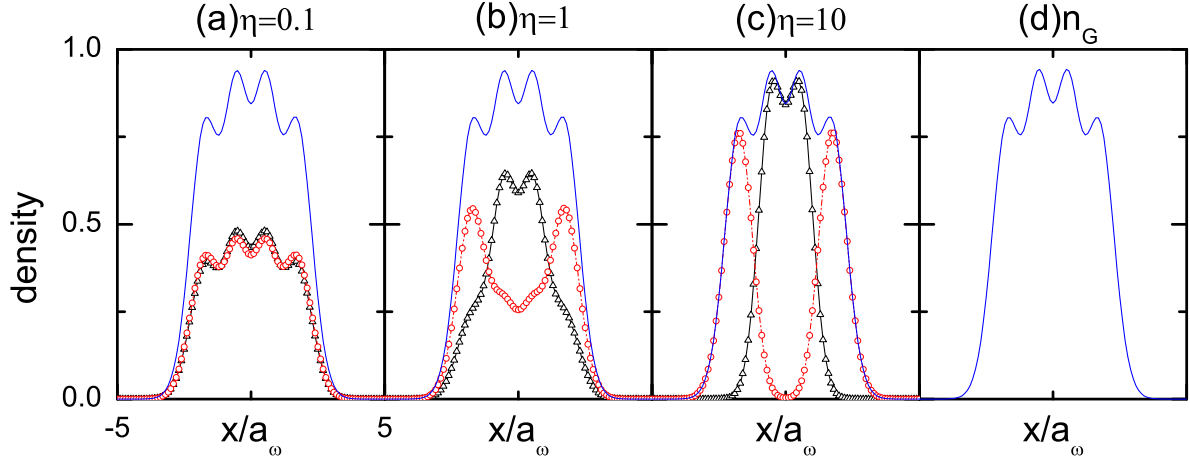}
\caption{(Color online) Ground state density distributions for the $2B2F$ system with various anisotropic interaction ratios: (a) $\eta=0.1$; (b) $\eta=1$; (c) $\eta=10$ calculated by the variational perturbation method. The red circle, black triangle and blue solid curve represent fermion, boson and total density distributions, respectively. (d) Total density distribution in the TG limit.}
\end{figure}

It is interesting to indicate that at the limit $g_{bb} \rightarrow \infty$ but with finite $g_{bf}$, corresponding to the case of $\eta=0$, the system becomes a mixture of hard-core bosons and fermions \cite{Chen-BF}. In this case, the system can be mapped to a spin-1/2 Fermi gas by a generalized Bose-Fermi transformation:
\begin{eqnarray}
\Psi_b(x) &=&\exp \left[ i\pi \int_{-\infty }^x n_{\uparrow} \left(
z\right) dz\right] \Psi_{f \uparrow} \left( x\right) , \\
\Psi_f(x) &=&\exp \left[ i\pi \int_{-\infty }^\infty n_{\uparrow}
\left( z\right) dz\right]  \Psi _{f \downarrow} \left( x\right) , \label{JW2}
\end{eqnarray}
where $n_{\sigma}(x)=\Psi_{f,\sigma}^{\dagger}\left( x\right) \Psi_{f,\sigma}\left( x\right) $, and $\Psi_{f, \sigma}^\dagger \left( x\right) $ ( $\Psi_{f,\sigma}\left(x\right) $) is the creation (annihilation) operator at location
$x$ for $\sigma$-component fermions ($\sigma=\uparrow,\downarrow$). The second mapping in the above equations is introduced to enforce the
fermion operators $\Psi _{f,\uparrow}$ and $\Psi _{f,\downarrow}$ fulfilling the anti-commutation
relation $\{\Psi _{f,\uparrow}, \Psi _{f,\downarrow}\} = 0$. By this mapping, it is known that the density distributions of the Bose-Fermi mixture in the limit of $g_{bb} \rightarrow \infty$ are identical to the distributions of the corresponding spin-1/2 Fermi gas, and thus we have $n_b = n_f = n_{tot}/2$  ($n_{tot}=n_b + n_f $) when $N_b=N_f$ from the symmetry requirement of the exchange invariance for $\uparrow$ and $\downarrow$ fermions. From the above analysis, it is not hard to understand why we have $n_b \approx n_f$ for the equal-mixing system with $\eta \ll 1$ as shown in Fig.3 (a). Also the density distributions shown in Fig.2 (a) for the $2B1F$ system with $\eta \ll 1$ are consistent with the distributions of spin-1/2 Fermi gas in Ref. \cite{Guan09}, according to the above mapping in the limit of $\eta=0$.

\section{Summary and outlooks}
In summary, we have studied the properties of 1D Bose-Fermi mixtures at the strongly repulsively limit. For the exactly solvable model with equal boson-boson and boson-fermion interactions, we give the analytical expression for the ground state energy in the strongly interacting regime, which clearly indicates that the ground state energy is dependent on the Bose-Fermi configuration and the degeneracy in the infinitely repulsive limit is lifted when the interaction deviates this limit. For the general case with different boson-boson and boson-fermion interactions, we derive the universal energy relation of the mixture system and then study the few-particle systems in harmonic traps by the variational perturbation theory within the degenerate ground state subspace of the system in the infinitely repulsive limit. Our results show that the total ground-state density profile in the strongly repulsive regime is not sensitive to the anisotropy of interactions and Bose-Fermi configurations, which however have significant effects on the density distributions of bosons and fermions. The species-dependent density distributions may be experimentally detected in a similar way as in the recent experiment for a 1D two-component fermionc system \cite{Murmann}.

Our variational perturbation method can be directly extended to deal with larger systems with more particle numbers though the integral coefficients, which are closely related to Tan's contacts, become more complicated with the increase of particle number.
After constructing the variational basis, we can investigate the crossover from few-body to many-body systems in the same scheme described in the present paper. However, when the particle number is large, it becomes a very difficult task to determine the integral coefficients via carrying numerical multiple integral. For the large-$N$ system, it is more convenient to study the ground state properties by developing some methods based on the density functional theory \cite{DFT-1d,GaoXL,DFT-BB,DFT-BF}. It is also interesting to study the perturbation correction of the ground state energy for the large-N system in a harmonic trap by generalizing the method for the single-component bosonic system in Ref.\cite{Korepin-pra}. Our method can be also applied to study other multi-component systems with more inner degrees of freedom, for example spinor quantum gases with $S=1$ and $S=3/2$. Based on the variational perturbation method, we can also study the quantum magnetism for strongly interacting multi-component quantum gases in the future work.

\begin{acknowledgments}
The work is supported by NSFC under Grants No. 11425419, No. 11374354 and No. 11421092.
\end{acknowledgments}

\end{document}